\documentclass[11pt,english,reqno]{amsart}
\usepackage[T1]{fontenc}
\usepackage[latin1]{inputenc}
\usepackage[dvips]{graphicx}
\usepackage{amssymb}
\usepackage{amsfonts}
\usepackage{float}
\usepackage{babel}

\makeatletter

\theoremstyle{plain}
\newtheorem{cor}{Corollary}
\newtheorem{thm}{Theorem}

\newtheorem{lem}{Lema}

\newtheorem{Proper}{Property}
\theoremstyle{remark}

\theoremstyle{definition}
\newtheorem{defi}{Definition}
\newtheorem{ex}{Example}
\newtheorem{Note}{Remark}

\newcommand{\rz}{\mathbb{R}}

\newcommand{\N}{\mathbb{N}}

\makeatother

\begin{document}

\title{THE UNIVERSAL CARDINAL ORDERING OF FIXED POINTS }

\author{Jesús San Martín$^{a,b}$, Mª Jose Moscoso$^{a}$, A. González Gómez$^{c}$}


\maketitle
{\small $a$ Departamento de Matemática Aplicada, E.U.I.T.I. Universidad
Politécnica de Madrid. 28012-Madrid, SPAIN.}{\small \par}

{\small $b$ Departamento de Física Matemática y de Fluidos, Facultad
de Ciencias. Universidad Nacional de Educación a Distancia. 28040-Madrid,
SPAIN.}{\small \par}

{\small $c$ Departamento de Matemática Aplicada a los Recursos Naturales,
E.T. Superior de Ingenieros de Montes. Universidad Politécnica de
Madrid. 28040-Madrid, SPAIN.}{\small \par}

{\small Corresponding author:} {\small Jesús San Martín. e-mail:
jsm@dfmf.uned.es}{\small \par}

\begin{abstract}
We present the theorem which determines, by a permutation, the cardinal
ordering of fixed points for any orbit of a period doubling cascade.
The inverse permutation generates the orbit and the symbolic sequence
of the orbit is obtained as a corollary. The problem present in the
symbolic sequences is solved. There, repeated symbols appear, for
example, the R (right), which cannot be distinguished among them as
it is not known which R is the rightmost of them all. Therefore, there
is a lack of information about the dynamical system. Interestingly
enough, it is important to point that this theorem needs no previous
information about any other orbit.
\end{abstract}

\section{Introduction}

The period doubling cascade ~\cite{Fei1,Fei2} is one of the most
common phenomena in dynamical systems and appears as one of the most
diverse fields of investigation. As time goes on, it has been found
in physical \cite{lete}, chemical \cite{kiss}, and biological \cite{giacomo}
models, to name but a few. Despite it being ubiquitous and to be perhaps
the most important mechanism of transition to chaos, it has not been
completely characterized yet. Its importance and the lack of a complete
comprehension of this phenomenon influences negatively the adecuate
comprehension of all of the underlying phenomena, which as has been
said above, is extraordinarily widespread. A common way to characterize
an orbit of a period doubling cascade, generated by iteration of an
unimodal map with a critical point in C, is giving a dynamical sequence.
So, the 8-periodic orbit has the symbolic sequence orbit CRLRRRLR,
where R and L indicate, respectively, right and left with respect
to the critical point C. Immediately, a problem can be seen: some
R cannot be distinguished from others, and the same happens with the
L. The R situated in the fifth position of the symbolic sequence,
is more to the right or to the left than the R in the sixth position?
The only thing that is known is that some points are to the left side
and others to the right of C, but where exactly are those points?
Obviously, this lack of information makes it difficult or impossible
to perform any calculation. There is another problem, symbolic sequences
are obtained recursively from the sequence of a former orbit of the
period doubling cascade; therefore, the symbolic sequence of $2^{k}$-periodic
orbit ($k$ arbitrary) is not known, so it cannot be used to prove
new theorems in dynamical systems. In this paper we will provide solutions
to these problems.

Let us place this problem in its historical context, in order to see
its origin, evolution and pending issues. The classical work of Metropolis,
Stein and Stein ~\cite{metro} set the big basement of the study
of dynamical systems, demonstrating the universal behavior that emerged
from the unimodal map. In that pioneering work, it was demonstrated
that the transformations of an interval in itself generate limit sets
that have an universal structure. That is to say, they are independent
from the function itself and the only important thing is the function
to be unimodal. The limit sets are characterized by finite sequences
of the symbols R and L, depending on wether the successive iterates
are plotted right or left of the critical point of the unimodal function.This
finite sequence of symbols, the symbolic sequence, is repeated periodically
and represents a periodic orbit of the dynamical system. The underlying
universal behavior forces the ordering of the different orbits with
respect to their parameter value and also depending of their period.
This universal ordering was called to the attention of Feigenbaum.
In his famous works ~\cite{Fei1,Fei2} he found that the periodic
orbits duplicate their period according to a universal rate, as a
universal law, independently of the iterated function. This is the
well known Feigenbaum cascade or period doubling cascade. What originally
was a universal behavior, reflected as a symbolic sequence, was converted
into a numerical universal behavior allowing the forecast of the successive
bifurcations. The numerical and predictive character has converted
it into a powerful tool that has extended to continuous dynamical
systems. To reach this point it is necessary to build a return map.
If the return map is unimodal, then the dyanamical system will probably
show the universal behavior forecasted by Feigenbaum, and it will
be possible to determine where the successive bifurcations will appear.
Feigenbaum work is condensed into the Feigenbaum-Çvitanovic equation.
He conjectured the existence of only one relevant eigenvalue for this
equation. This eigenvalue would be responsible for the universal behavior
found by Feigenbaum. Later, these hypothesis got a rigorous mathematical
proof \cite{col}. Within this body of knowledge, there are two relevant
problems closely related that emerge from the Feigenbaum cascade:

i) Given the symbolic sequence of a given periodic orbit, it is possible
to get the symbolic sequence of the doubled-period orbit. This is
a recursive process that necessarily requires the knowledge of the
former element. Therefore, it is not possible to know directly the
symbolic sequence of an arbitrary orbit and it is not possible to
use it in mathematical proofs. This rises the question of how to directly
determine the symbolic sequence of an arbitrary orbit without calculating
the previous sequences. 

ii) Related with the former point, another question arises. The symbolic
sequence indicates that the successive iterates are to the right (R)
or to the left (L) of the critical point of the iterated unimodal
function. But, how far to the right or to the left are these found
in relation to the others? Which is their relative position? Or more
rigorously stated: should iterates be enumerated, what their cardinals
would be?

First, we will obtain the solution to the second problem, obviously
much more complicated; but a simple corollary, which is easy and fast
to prove, will provide the solution to the first problem.

Observe that the points of the $2^{k}$-periodic orbit are fixed points
of $f^{2^{k}}$. The stability and bifurcations of the fixed points
and the flow in their neighborhoods play an important role in dynamical
systems theory. Positions of fixed points of mappings are not only
useful in nonlinear dynamics, they are also important in quamtum mechanics
~\cite{and} and therefore in all related subjects. Hence, we expect
the theorems we are going to prove will be of use in these subjects.

This paper is organized as follows. First, the definitions and notations
necessary to prove the theorems will be introduced. Later, a first
result will be obtained that establishes the relative position of
the points of the $2^{k+1}$-periodic orbit in function of the relative
position of the points of the $2^{k}$-periodic orbit. This result
will be used to prove a theorem that determines the cardinal ordering
of the points in the $2^{k+1}$-periodic orbit in terms of the cardinal
position of the points in the $2^{k}$-periodic orbit. Based on these
preceding results, the final goal is obtained: the theorem that determines
the cardinal ordering of the points in the $2^{k}$-periodic orbit,
for an arbitrary $k$, with no previous information. Corollaries and
reformulations of the theorems will be stated along the process. Examples
will also be shown to understand the geometrical meaning of the results
and to ease the use of the theorems by scientists and engineers.

\medskip{}

\section{Definitions and notation}

Let $f:\mathrm{I}\subset\rz\to\mathrm{I}$ be an unimodal map,
where $C$ denotes its critical point. Let $O=\{ x_{0},x_{1},\ldots,x_{q-1}\}$
be a $q$-periodic orbit of $f$, where $x_{0}=C$, then the first
and the second iterates of $C$ determine the subinterval J$=[f^{2}(C),f(C)]$
when $f$ has a maximum, (J$=[f(C),f^{2}(C)]$ if $f$ has a minimum)
such that $O\subset[f^{2}(C),f(C)]\subset\mathrm{I}.$

Given that the orbit $O$ does not have the natural order within the
interval J$=[f^{2}(C),f(C)]$ we introduce the following definition.

\begin{defi} The set $\{ C_{(1,q)}^{*},\, C_{(2,q)}^{*},\ldots,C_{(q,q)}^{*}\}$
will denote the ascending (descending) cardinality order of the orbit
$O=\{ C,f(C),\ldots,f^{q-1}(C)\}$ when $f$ has a minimum (maximum)
in $C$. Given that $C_{(i,q)}^{*}\ i=1,\ldots,q$, is the point of
the orbit $O$ plot in the cardinal position $'i'$ when this orbit
is cardinally ordered, the point $C_{(i,q)}^{*}$ is defined as the
$i-$th cardinal of the $q-$period orbit. \end{defi} \begin{Note}\label{not1}
Notice that $f(C)=C_{(1,q)}^{*}$ and $f^{2}(C)=C_{(q,q)}^{*}.$ Therefore,
if $f$ has a maximum in $C$, it results in

\[
f^{2}(C)=C_{(q,q)}^{*}<\ldots<C_{(2,q)}^{*}<C_{(1,q)}^{*}=f(C);\]
 meanwhile that if $f$ has a minimum in $C$ , it results in 

\[
f(C)=C_{(1,q)}^{*}<\ldots<C_{(2,q)}^{*}<C_{(q,q)}^{*}=f^{2}(C)\]
 \end{Note}

\begin{defi}\label{def2} The natural number ${\sigma}_{(i,q)},\ i=1,\ldots,q$
\, will denote the number of iterations of $f$ such as that $f^{\sigma_{(i,q)}}(C)=C_{(i,q)}^{*},\ i=1,\ldots,q$.
\end{defi}

The goal of this paper is to determine the number of iterations, $\sigma_{(i,q)},$
of $f$ from $C$, to get the point of the $q$-periodic orbit ($q=2^{p}$)
situated in the $i-$th cardinal position $C_{(i,q)}^{*}$.

\begin{defi} We denote as {\Large {${\sigma}_{q}$}} the permutation
{\Large $\sigma_{q}$}$=(\sigma_{(1,q)},\sigma_{(2,q)},\ldots,\sigma_{(q,q)})$,
that is the $q$-tuple formed by the ${\sigma}_{(i,q)}$. \end{defi}

\begin{defi} Let $\{ a_{1},a_{2},\ldots,a_{n}\}$ be a sequence of
real numbers. We define the \emph{Reflection} of the sequence $\{ a_{1},a_{2},\ldots,a_{n}\}$
with increment $\alpha$, denoted as $R(a_{1},\cdots,a_{n};\alpha),$
as the sequence of real numbers given by:

\[
R(a_{1},\cdots,a_{n};\alpha)=\{ a_{1},\cdots,a_{n},a_{n}+\alpha,a_{n-1}+\alpha,\ldots,a_{1}+\alpha\}.\]
 \end{defi}

\begin{defi}\label{def5} Let $\{ a_{1},a_{2},\ldots,a_{n}\}$ be
a sequence of real numbers. We define the \emph{Iterated Reflection}
of the sequence $\{ a_{1},a_{2},\ldots,a_{n}\}$ with the descending
increment $2^{k}$, $k\in\N$, and we denote it as $\partial R(a_{1},\cdots,a_{n};2^{k})$,
as the sequence of real numbers given by:

\[
\partial R(a_{1},\cdots,a_{n};2^{k})=R(R(\ldots(R(R(a_{1},\cdots,a_{n});2^{k});2^{k-1})\ldots;2^{1});2^{0})\]
 \end{defi} \begin{Proper}\label{pro} The Reflection verifies that\[
R(a_{1}+b_{1},a_{2}+b_{2},\ldots,a_{n}+b_{n};\alpha+\beta)=R(a_{1},a_{2},\ldots,a_{n}+;\alpha)+R(b_{1},b_{2},\ldots,+b_{n};\beta)\]
 \end{Proper}

\section{Theorems}

\begin{lem}
\label{odd} Let $f:\mathrm{I}\to\mathrm{I}$ be an unimodal map,
depending on a parameter, that undergoes a period-doubling cascade.
Let $\{ C,f(C)=C'\}$ be $2-$periodic superstable orbit of the cascade,
where $C$ is the critical point of $f$. Let $O'\equiv\{ C,y_{1},\cdots,y_{2^{p}-1}\}$
and $O\equiv\{ C,x_{1},\cdots,x_{2^{p+1}-1}\}$ be the $2^{p}-$periodic
and $2^{p+1}-$periodic superstable orbits of $f$, respectively.
Then
\begin{enumerate}
\item [(i)] The relative position of the points $O_{\text{odd}}\equiv\{ x_{1},x_{3},x_{5},\ldots x_{2^{p+1}-3},C'\}$
with respect to $C'$ is the same as the relative position of the
points $O'\equiv\{ C,y_{1},\ldots,y_{2^{p}-1}\}$ with respect to
$C$. The points are visited in the same order in both cases. 
\item [(ii)] The relative position of the points $O_{\text{even}}\equiv\{ C,x_{2},x_{4},\ldots,x_{2^{p+1}-2}\}$
with respect to $C$ is the same as the relative postion of the points
of the orbit $O'\equiv\{ C,y_{1},\ldots,y_{2^{p}-1}\}$ with respect
to $C$ after being conjugated by homotecy with respect to $C.$ 
\end{enumerate}
\end{lem}
\begin{proof}
Let us suppose, without loss of generality, that $f$ has a maximum
at the critical point $C.$ Therefore, $f^{2}$ will have a minimum
at the critical point $C$ and a maximum at $C'.$ Consequently, there
is a neighborhood of $C,$ $\mathrm{I}_{C}$, and a neighborhood of
$C',$ $\mathrm{I}_{C'}$, such that $f_{|_{\mathrm{I}_{C}}}^{2}$
and $f_{|_{\mathrm{I}_{C'}}}^{2}$ are unimodals (see fig. \ref{fig:1a-and-1b}a).%
\begin{figure}
\begin{tabular}{cc}
\includegraphics[width=0.45\textwidth]{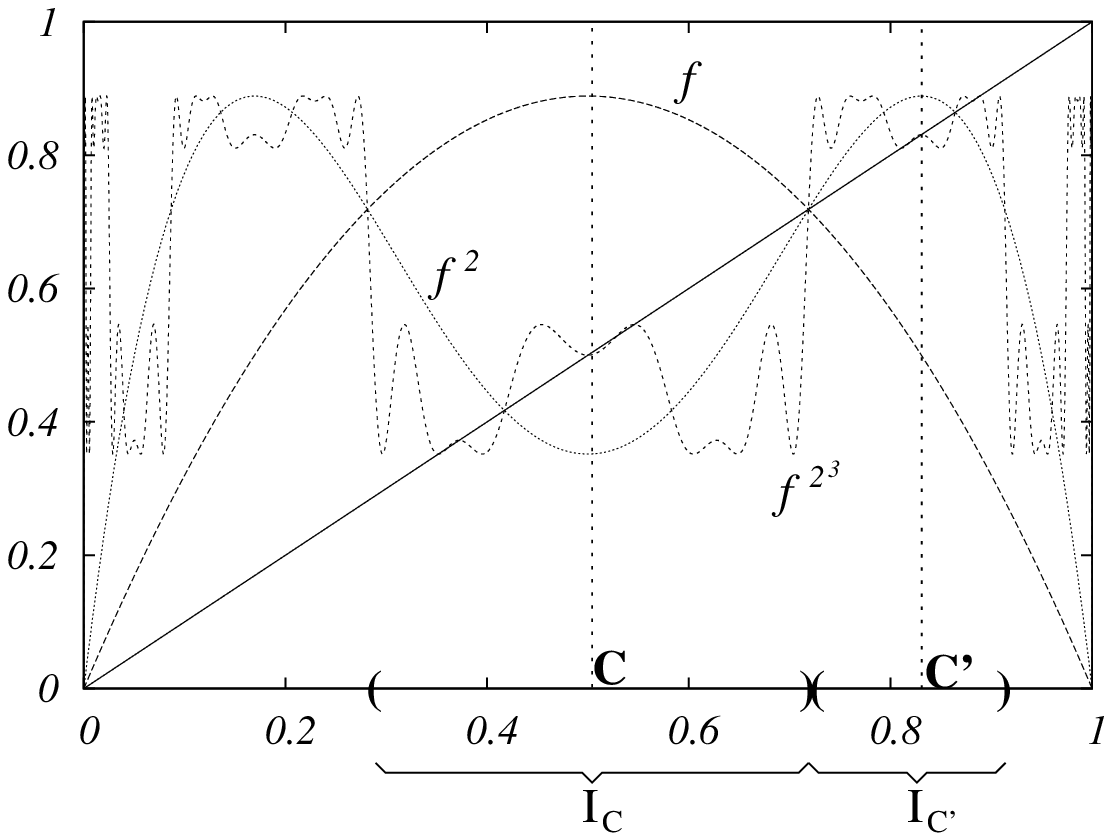}&
\includegraphics[width=0.45\textwidth]{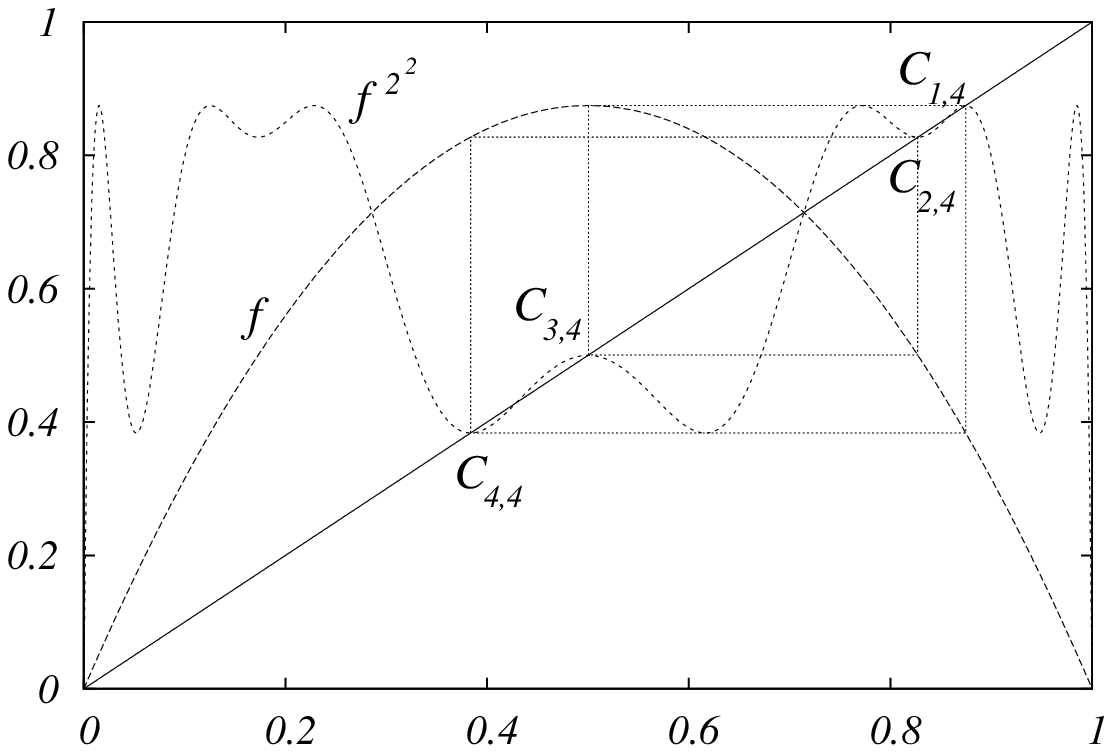}\\
(a)&
(b)\\
\end{tabular}

\caption{\label{fig:1a-and-1b}(a) $f$ is an unimodal map, and $f^{2}$is
also unimodal in the neighbourhood of $I_{C}$ and $I_{C^{\prime}}$.
The graphs of $f^{2^{2}}$are reproduced in the neighbourhood of $I_{C}$
and $I_{C^{\prime}}$, being conjugated of each other. (b) Graph of
$f^{2^{2}}$.}
\end{figure}

As $O\equiv\{ C,x_{1},\ldots,x_{2^{p+1}-1}\}$ is the $2^{p+1}-$periodic
superstable orbit of $f$, it turns out that $O_{even}\equiv\{ C,x_{2},x_{4},\ldots,x_{2^{p+1}-2}\}$
and $O_{odd}\equiv\{ x_{1},x_{3},\ldots,x_{2^{p+1}-1}\}$ are the
$2^{p}-$periodic superstable orbits of $f_{|_{\mathrm{I}_{C}}}^{2}$
and $f_{|_{\mathrm{I}_{C'}}}^{2}$ respectively.

As $f$ undergoes a period-doubling cascade so do $f_{|_{\mathrm{I}_{C}}}^{2}$
and $f_{|_{\mathrm{I}_{C'}}}^{2}$ and viceversa. Therefore, when
$f_{|_{\mathrm{I}_{C}}}^{2}$ goes through a bifurcation and it goes
from having a $2^{p-1}-$periodic superstable orbit to having a $2^{p}-$periodic
superstable orbit, then $f$ goes from having a $2^{p}-$periodic
superstable orbit to having a $2^{p+1}-$periodic superstable orbit.
Furthermore, $f_{|_{\mathrm{I}_{C'}}}^{2}$ and $f_{|_{\mathrm{I}_{C}}}^{2}$
undergo bifurcations simultaneously.

It is well known that if two unimodal maps, with a maximum at their
critical points, undergo a period double cascade, then they will have
the same symbolic sequence. Meanwhile, if one of them has a maximum
and the other a minimum, those symbolic sequences will be conjugated
\cite{Gil}.

Observe that as $f_{|_{\mathrm{I}_{C'}}}^{2}$ is unimodal at $\mathrm{I}_{C'}$,
if we take $x=x_{2^{p+1}-1}$ then \[
(f^{2}(x_{2^{p+1}-1}))'=(f(\underbrace{f(x_{2^{p+1}-1})}_{C}))'=f'(C)f'(x_{2^{p+1}-1})=0.\]
 Therefore, $x_{2^{p+1}-1}$ is an extremum of $f_{|_{\mathrm{I}_{C'}}}^{2}$
and since $f_{|_{\mathrm{I}_{C'}}}^{2}$ is unimodal, and thus it
has one single extremum at $C'$, it results in $C'=x_{2^{p+1}-1}$.

In line with the former discussion:
\begin{enumerate}
\item [(i)] follows because $f$ and $f_{|_{\mathrm{I}_{C'}}}^{2}$ are
unimodal, with maxima at $C$ and $C'$, respectively. 
\item [(ii)] is also deduced since $f$ and $f_{|_{\mathrm{I}_{C}}}^{2}$
are unimodal with a maximum and minimum at $C$, respectively. The
conjugation of the orbit $O'\equiv\{ C,y_{1},\ldots,y_{2^{p}-1}\}$
with respect to $C$, is geometrically a homotecy of $O'$ with respect
to $C$; it is the equivalent to the homotecy of $O_{odd}$ with respect
to $C'$ and then a translation to $C$.

(See figures \ref{fig:1a-and-1b} and \ref{fig:2a-and-2b}).

\end{enumerate}
\end{proof}
\begin{figure}
\begin{tabular}{cc}
\includegraphics[width=0.45\textwidth]{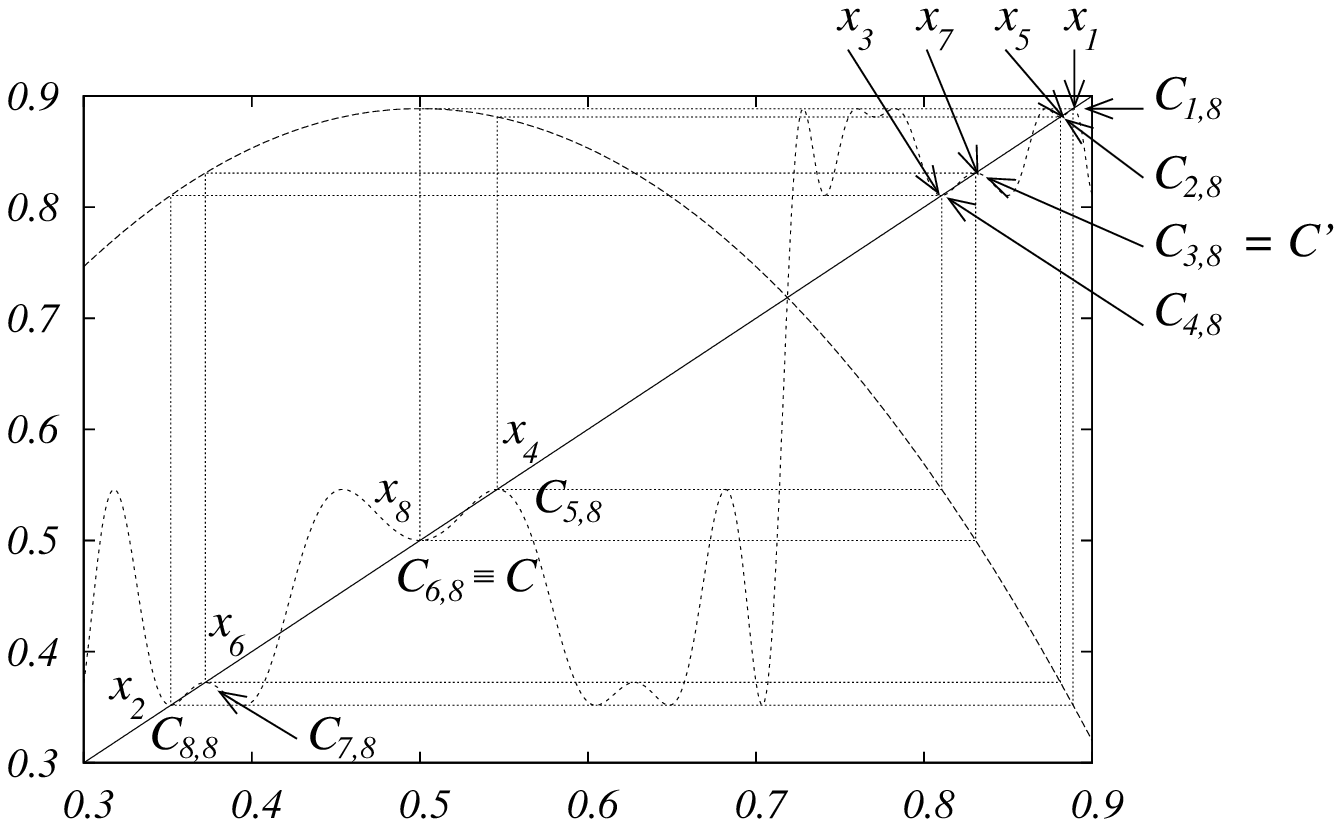}&
\includegraphics[width=0.45\textwidth]{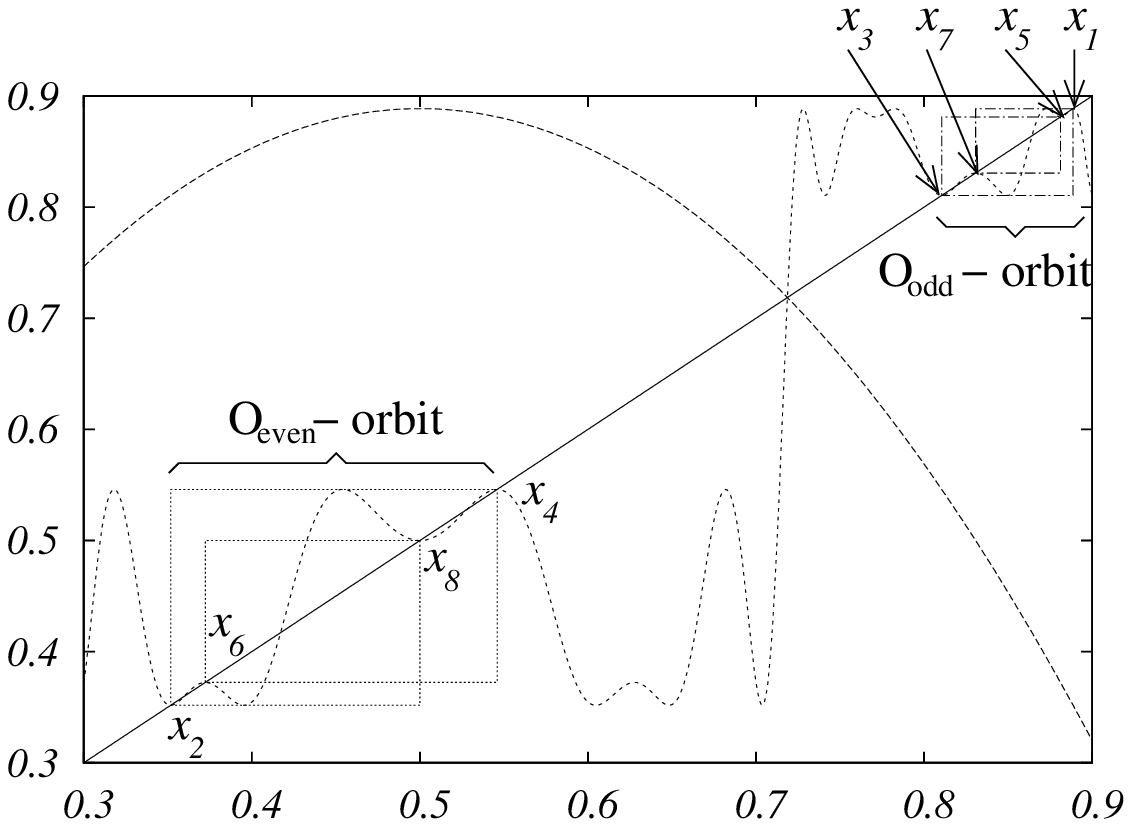}\\
(a)&
(b)\\
\end{tabular}

\caption{\label{fig:2a-and-2b}(a) Cardinals of the points of the $2^{3}$-periodic
superstable orbit (b) $O_{even}$ reproduces $O_{odd}$ conjugated.
$O_{odd}$ is the same orbit observed in fig. \ref{fig:1a-and-1b}b.}
\end{figure}

\begin{lem}
\label{lem:tres-dos}Let $f:\mathrm{I}\to\mathrm{I}$ be an unimodal
map, depending on a parameter, that, as a function of the parameter,
goes through a period doubling cascade. If \[
C\,\mathrm{I}_{1}\,\mathrm{I}_{2}\ldots\mathrm{I}_{2^{p}-1}\quad\text{ where }\ \mathrm{I}_{i}=\mathrm{R}\,\text{ or }\,\mathrm{L}\quad i=1,\ldots,2^{p}-1\]
 is the symbolic sequence of the $2^{p}-$periodic superstable orbit
of the cascade, then \[
C\,\mathrm{R}\,\overline{\mathrm{I}}_{1}\,\mathrm{R}\,\overline{\mathrm{I}}_{2}\,\mathrm{R}\ldots\mathrm{R}\,\overline{\mathrm{I}}_{2^{p}-1}\mathrm{R}\]
is the symbolic sequence of the $2^{p+1}-$periodic superstable orbit.
Where

\[
\overline{\mathrm{I}}_{i}=\left\{ \begin{array}{ll}
\mathrm{R} & \hbox{if}\ \mathrm{I}_{i}=\mathrm{L}\\
\mathrm{L} & \hbox{if}\ \mathrm{I}_{i}=\mathrm{R}\end{array}\right.\qquad\, i=1,\ldots,2^{p}-1.\]
 
\end{lem}
\begin{proof}
The points of the $2^{p+1}-$periodic superstable orbit $O\equiv\{ C,x_{1},\cdots,x_{2^{p+1}-1}\}$
can be split into the orbits $O_{\text{even}}\equiv\{ C,x_{2},\cdots,x_{2^{p+1}-2}\}$
and $O_{\text{odd}}\equiv\{ x_{1},\cdots,x_{2^{p+1}-1}\}$. Where,
$O_{\text{even}}$ is the $2^{p}-$periodic superstable orbit of $f_{|_{\mathrm{I}_{C}}}^{2}$
and $O_{\text{odd}}$ is the $2^{p}-$periodic superstable orbit of
$f_{|_{\mathrm{I}_{C'}}}^{2}.$

By lemma, ~\ref{odd} the $2^{p}-$periodic superstable orbit of
$f,$ after being conjugated by homotecy with respect to $C$, is
transformed into $O_{\text{even}}$. As the homotecy with respect
to $C$ generates the conjugation $\mathrm{R}\longleftrightarrow\mathrm{L}$,
that is to say, it transforms $\mathrm{I}_{i}$ into $\overline{\mathrm{I}}_{i}\quad i=1,\ldots,2^{p}-1$,
and as the symbolic sequence of the $2^{p}-$periodic superstable
orbit is \[
C\,\mathrm{I}_{1}\,\mathrm{I}_{2}\ldots\mathrm{I}_{2^{p}-1},\]
it results that the symbolic sequence of $O_{\text{even}}$ is

\[
C\,\mathrm{\overline{I}}_{1}\,\mathrm{\overline{I}}_{2}\ldots\mathrm{\overline{I}}_{2^{p}-1}.\]

On the other hand, all of the points of $O_{\text{odd}}$ are in the
neighborhood of $\mathrm{I}_{C'}$, situated to the right of $C$;
therefore, each time that a point of $O_{\text{odd}}$ is visited,
it will appear as a $\mathrm{R}$ in the sequence of $O$. Given that
the orbit of $O$ is built by alternating points of $O_{\text{odd}}$
and $O_{\text{even}},$ if an $\mathrm{R}$ is put after every symbol
of the symbolic sequence of $O_{\text{even}}$ , then the symbolic
sequence of $O$ will be obtained. That is to say, as the symbolic
sequence of $O_{\text{even}}$ is \[
C\,\mathrm{\overline{I}}_{1}\,\mathrm{\overline{I}}_{2}\ldots\mathrm{\overline{I}}_{2^{p}-1}\]
 the sought for symbolic sequence results in \[
C\,\mathrm{R}\,\mathrm{\overline{I}}_{1}\,\mathrm{R}\,\mathrm{\overline{I}}_{2}\,\mathrm{R}\ldots\mathrm{R}\,\mathrm{\overline{I}}_{2^{p}-1}\,\mathrm{R}\]
 
\end{proof}
As a trivial case, if the superstable orbit of period $1$, with symbolic
sequence $C$, is taken it results that the sequence of the $2$-periodic
superstable orbit is $C\,\mathrm{R}$ as it is well-known.

\begin{ex}\label{pri1} While using the symbolic sequence of the
$2^{3}$-periodic superstable orbit, which is $C\,\mathrm{R}\,\mathrm{L}\,\mathrm{R}\,\mathrm{R}\,\mathrm{R}\,\mathrm{L}\,\mathrm{R},$
we are going to obtain the symbolic sequence of the $2^{4}$-periodic
superstable orbit using the technique of MSS \cite{metro} and later
the lemma ~\ref{lem:tres-dos}.

\begin{enumerate}
\item [i)] To get the symbolic sequence of the $2^{4}$-periodic superstable
orbit according to MSS, the following steps must be taken:

\begin{itemize}
\item [a)] To write consecutively twice the symbolic sequence: \[
C\,\mathrm{R\, L\, R\, R\, R\, L\, R}\, C\,\mathrm{R\, L\, R\, R\, R\, L\, R}\]
 
\item [b)] Change the second $C$ to a $\mathrm{R}$ if the $\mathrm{R-}$parity
of the orginal orbit is even or $\mathrm{L}$ otherwise. As in this
case the $\mathrm{R-}$parity is odd we have to change the second
$C$ for a $\mathrm{L}$ , resulting in \[
C\,\mathrm{R\, L\, R\, R\, R\, L\, R\,\mathbf{L}\, R\, L\, R\, R\, R\, L\, R}\]

\end{itemize}
\item [ii)] Using the Lemma~\ref{lem:tres-dos} to obtain the symbolic
sequence of the $2^{4}$-periodic superstable orbit, the following
steps must be taken: 

\begin{itemize}
\item [a)] Conjugate the symbolic sequence of the original orbit: \[
C\,\mathrm{R\, L\, R\, R\, R\, L\, R}\quad\Longrightarrow\quad C\,\mathrm{L\, R\, L\, L\, L\, R\, L}\]
 
\item [b)] To add a $\mathrm{R}$ after every symbol of the conjugated
sequence

\[
C\,\mathrm{R\, L\, R\, R\, R\, L\, R\, L\, R\, L\, R\, R\, R\, L\, R}\]
 obtaining the same results.

\end{itemize}
\end{enumerate}
\end{ex}

In what follows, we are going to use the Lemma ~\ref{odd} to determine
the cardinals of the points of the $2^{p+1}-$periodic superstable
orbit when the cardinals of the points of the $2^{p}-$periodic superstable
orbit are known.

\begin{thm}
\label{orbit} Let $f:\mathrm{I}\to\mathrm{I}$ be an unimodal map,
depending on a parameter, undergoing a period doubling cascade. Let
$\{ C_{(1,2^{p})}^{*},\, C_{(2,2^{p})}^{*},\ldots,C_{(2^{p},2^{p})}^{*}\}$
be the cardinals of the points of the $2^{p}$-periodic superstable
orbit of the cascade, given by $f^{{\sigma}_{(i,2^{p})}}(C)=C_{(i,2^{p})}^{*}$
$,\quad i=1,\ldots,2^{p}.$ Then the permutation {\Large $\sigma_{2^{p+1}}$},
given by \[
\text{{\Large{$\sigma_{2^{p+1}}$}}}=(\sigma_{(1,2^{p+1})},\ldots,\sigma_{(2^{p+1},2^{p+1})})=R(2{\sigma}_{(i,2^{p})}-1,\ldots,2{\sigma}_{(2^{p},2^{p})}-1;1)\]
 determines the cardinals $\{ C_{(1,2^{p+1})}^{*},\, C_{(2,2^{p+1})}^{*},\ldots,C_{(2^{p},2^{p+1})}^{*},\ldots,C_{(2^{p+1},2^{p+1})}^{*}\}$
of the points of the $2^{p+1}$-periodic superstable orbit by means
of $f^{\sigma_{(i,2^{p+1})}}(C)=C_{(i,2^{p+1})}^{*}$ $\quad i=1,\ldots,2^{p+1}.$
\end{thm}
\begin{proof}
Let $\{ C_{(1,2^{p+1})}^{*},\, C_{(2,2^{p+1})}^{*},\ldots,C_{(2^{p},2^{p+1})}^{*},C_{(2^{p}+1,2^{p+1})}^{*}\ldots,C_{(2^{p+1},2^{p+1})}^{*}\}$
be the cardinals of the $2^{p+1}$-periodic superstable orbit given
by $O\equiv\{ C,x_{1},x_{2},\ldots,x_{2^{p+1}-1}\}.$

By Lemma ~\ref{odd}, the $2^{p}$ first cardinals of $O$, given
by $\{ C_{(1,2^{p+1})}^{*},\ldots,C_{(2^{p},2^{p+1})}^{*}\}$, are
the cardinals of the orbit $O_{\text{odd}}$ of $O$ (See figures
\ref{fig:2a-and-2b}).

Furthermore, by Lemma ~\ref{odd}, they are visited in the same order
as the orbit of period $2^{p}$, with two nuances: 
\begin{enumerate}
\item [(a)] the points of $O_{\text{odd}}$ are reached by means of $f^{2}$. 
\item [(b)] the points of $O_{\text{odd}}$ are reached from the critical
point $C'$, and not from $C.$ 
\end{enumerate}
Therefore, this results in

\[
(f^{2})^{{\sigma}_{(i,2^{p})}}(C')=f^{2{\sigma}_{(i,2^{p})}}(C')=C_{(i,2^{p+1})}^{*}\qquad i=1,\ldots,2^{p}\]

Furthermore, as 

\[
f^{2^{p+1}-1}(C)=C'\]
results in \[
f^{2{\sigma}_{(i,2^{p})}}(C')=f^{2{\sigma}_{(i,2^{p})}}(f^{2^{p+1}-1}(C))=f^{2{\sigma}_{(i,2^{p})}-1}(C)\]
that is \[
f^{2{\sigma}_{(i,2^{p})}-1}(C)=C_{(i,2^{p+1})}^{*}\qquad i=1,\ldots,2^{p}\]

On the other hand, the cardinals $\{ C_{(2^{p}+1,2^{p+1})}^{*},\, C_{(2^{p}+2,2^{p+1})}^{*}\ldots,C_{(2^{p+1},2^{p+1})}^{*}\}$
are the cardinals of $O_{\text{even}}$(See figures \ref{fig:2a-and-2b}).
By the Lemma ~\ref{odd}, as it was shown in the proof, $O_{\text{even}}$
is obtained applying a homotecy to $O_{\text{odd}}$ with respect
to ${C}'$ and then translating to $C$. Therefore, as \[
O_{\text{odd}}=\{ f^{2{\sigma}_{(1,2^{p})}}(C'),f^{2{\sigma}_{(2,2^{p})}}(C'),\ldots,f^{2{\sigma}_{(2^{p},2^{p})}}(C')\}\]
applying the homotecy with respect to ${C}'$ results in \[
O_{\text{even}}=\{ f^{2{\sigma}_{(2^{p},2^{p})}}(C'),f^{2{\sigma}_{(2^{p}-1,2^{p})}}(C')\ldots,f^{2{\sigma}_{(1,2^{p})}}(C')\}\]

After translation from $C'$ to $C$ results in

\[
O_{\text{even}}=\{ f^{2{\sigma}_{(2^{p},2^{p})}}(C),f^{2{\sigma}_{(2^{p}-1,2^{p})}}(C)\ldots,f^{2{\sigma}_{(1,2^{p})}}(C)\}\]

Joining the cardinals of $O_{\text{odd}}$ and $O_{\text{even}}$
to retrieve the $2^{p+1}$-periodic orbit finally results in \begin{multline*}
\{ C_{(1,2^{p+1})}^{*},\ldots,C_{(2^{p},2^{p+1})}^{*},C_{(2^{p}+1,2^{p+1})}^{*},\ldots,C_{(2^{p+1},2^{p+1})}^{*}\}=\\
=\left\{ \, f^{2{\sigma}_{(1,2^{p})}-1}(C),\ldots,f^{2{\sigma}_{(2^{p},2^{p})}-1}(C),\, f^{2{\sigma}_{(2^{p},2^{p})}}(C),\ldots,f^{2{\sigma}_{(1,2^{p})}}(C)\right\} \end{multline*}
that is \[
f^{{\sigma}_{(i,2^{p+1})}}(C)=C_{(i,2^{p+1})}^{*}\,,\qquad i=1,\ldots,2^{p+1}\]
 with \begin{multline*}
\text{\Large{$\sigma_{2^{p+1}}$}}=(\sigma_{(1,2^{p+1})},\ldots,\sigma_{(2^{p+1},2^{p+1})})=\\
=(2{\sigma}_{(1,2^{p})}-1,2{\sigma}_{(2,2^{p})}-1,\ldots,2{\sigma}_{(2^{p},2^{p})}-1,2{\sigma}_{(2^{p},2^{p})},\ldots,2{\sigma}_{(1,2^{p})})=\\
=R(2{\sigma}_{(1,2^{p})}-1,\ldots,2{\sigma}_{(2^{p},2^{p})}-1;1)\end{multline*}
 
\end{proof}
\begin{ex}\label{prim} Observing the $2^{2}$-periodic orbit in
fig. \ref{fig:1a-and-1b}b results that \[
\sigma_{2^{2}}=\begin{pmatrix}1 & 2 & 3 & 4\\
1 & 3 & 4 & 2\end{pmatrix}\ \begin{tabular}{c}
 $\longleftarrow$ \text{indicates the position 'i' of $C_{{(i,2^{2})}}^{*}$} \\
$\longleftarrow\ \text{number of interations},\sigma_{(i,2^{2})},\text{to reach \mbox{$C_{{(i,2^{2})}}^{*}$}}$ \end{tabular}{\Longrightarrow}\ f^{\sigma_{(i,2^{2})}}(C)=C_{(i,2^{2})}^{*}.\]
 Using the theorem \ref{orbit} and having in mind {\Large $\sigma_{2^{2}}$},
it is obtained 

\[
\sigma_{2^{3}}=\begin{pmatrix}1 & 2 & 3 & 4 & 5 & 6 & 7 & 8\\
1 & 5 & 7 & 3 & 4 & 8 & 6 & 2\end{pmatrix}\ \begin{tabular}{c}
 $\longleftarrow$ \text{ position 'i' of $C_{{(i,2^{3})}}^{*}$} \\
$\longleftarrow\ \text{ iterations},\sigma_{(i,2^{3})},\text{ to reach\mbox{\text{ $C_{{(i,2^{3})}}^{*}$}} }$ \end{tabular}{\Longrightarrow}\ f^{\sigma_{(i,2^{3})}}(C)=C_{(i,2^{3})}^{*}\]
 that, it can be seen, is the permutation obtained from the ${2^{3}}$-periodic
orbit in fig{\Large .} \ref{fig:2a-and-2b}a. \end{ex}

~

Finally, with the following theorem, we will reach our original goal
of determining how many iterations of $f$ over $C$ are necessary
to reach the point of the $2^{p}$-periodic orbit situated at the
cardinal position $'i'$, denoted by $C_{(i,2^{p})}^{*},\quad i=1,\ldots,2^{p}.$
The number of iterations needed to reach these points will be determined
by a permutation, whose inverse will determine the orbit and its symbolic
sequence. 

Notice that the orbit is arbitrary and no previous information of
any other orbit is needed, in contrast with the previous theorem where
the former orbit cardinals were needed.

\begin{thm}
\label{fuerte} Let $f:\mathrm{I}\to\mathrm{I}$ be an unimodal function
undergoing a period doubling cascade and the $2^{p}$-periodic orbit
of the cascade with cardinals $\{ C_{(1,2^{p})}^{*},\ldots,C_{(2^{p},2^{p})}^{*}\}$.
Let the permutation {\Large $\sigma_{2^{p}}$}$=(\sigma_{(1,2^{p})},\,\sigma_{(2,2^{p})},\ldots,\sigma_{(2^{p},2^{p})})\qquad p\in\N,\quad p\geq2$
be determined by\[
\text{{\Large{$\sigma_{2^{p}}$}}}=\partial R(1,2^{p-1}+1;2^{p-2})\qquad p\in\N,\quad p\geq2,\]
 then \[
f^{\sigma_{(i,2^{p})}}(C)=C_{(i,2^{p})}^{*}\qquad\, i=1,\ldots,2^{p}.\]
 
\end{thm}
\begin{proof}
From remark \ref{not1} the $2-$periodic orbit satisfies \[
f^{1}(C)=C_{(1,2)}^{*}\quad\text{ and }\quad f^{2}(C)=C_{(2,2)}^{*}\]
Using the definition \ref{def2} it results in \[
C_{(1,2)}^{*}=f^{\sigma_{(1,2)}}(C)\quad\text{ and }\quad C_{(2,2)}^{*}=f^{\sigma_{(2,2)}}(C)\]
 therefore, it results in \[
(\sigma_{(1,2)},\sigma_{(2,2)})=(1,2).\]
 We are going to prove the formula by induction over $p$ 
\begin{itemize}
\item [(i)] Case $p=2.$ The orbit of period $2^{2}$ is $\{ C,f(C),f^{2}(C),f^{3}(C)\}\equiv\{ C,\, C_{(1,2^{2})}^{*}\, C_{(2^{2},2^{2})}^{*},f^{3}(C)\}$,
where remark \ref{not1} has been used. Since the orbit of period
$2$ is given by $(\sigma_{(1,2)},\sigma_{(2,2)})=(1,2)$, applying
the Theorem ~\ref{orbit} results in \begin{multline*}
\text{{\Large{$\sigma_{2^{2}}$}}}=(\sigma_{(1,2^{2})},\sigma_{(2,2^{2})},\sigma_{(3,2^{2})},\sigma_{(2^{2},2^{2})})=R(2\sigma_{(1,2)}-1,2\sigma_{(2,2)}-1;1)=(1,3,4,2)=\\
=\partial R(1,3;1)=\partial R(1,2^{2-1}+1;2^{0}).\end{multline*}
 
\item [(ii)] Let us suppose that the theorem is valid for $p=k$ and check
wether this is true for $p=k+1,$ that is to say that \[
(\sigma_{(1,2^{k+1})},\sigma_{(2,2^{k+1})},\ldots,\sigma_{(2^{k+1},2^{k+1})})=\partial R(1,2^{k}+1;2^{k-1}).\]

Using the theorem ~\ref{orbit} and the property \ref{pro} we have
\begin{multline}
(\sigma_{(1,2^{k+1})},\sigma_{(2,2^{k+1})},\ldots,\sigma_{(2^{k+1},2^{k+1})})=R(2\sigma_{(1,2^{k})}-1,\ldots,2\sigma_{(2^{k},2^{k})}-1;1)=\\
=2R(\sigma_{(1,2^{k})},\sigma_{(2,2^{k})},\ldots,\sigma_{(2^{k},2^{k})};0)+R(\underbrace{-1,\ldots,-1}_{2^{k}};1).\end{multline}
By the induction hypothesis, and using recursively the property \ref{pro},
the equality $(1)$ is transformed into \begin{multline}
(\sigma_{(1,2^{k+1})},\sigma_{(2,2^{k+1})},\ldots,\sigma_{(2^{k+1},2^{k+1})})=2R(\partial R(1,2^{k-1}+1;2^{k-2});0)+R(\underbrace{-1,\ldots,-1}_{2^{k}};1)=\\
=2R\underbrace{(R(\cdots R(R}_{k-1}(1,2^{k-1}+1;2^{k-2});2^{k-3})\cdots;1);0)+R(\underbrace{-1,\ldots,-1}_{2^{k}};1)=\\
=R\underbrace{(R(\cdots R(R}_{k-1}(2,2(2^{k-1}+1);2^{k-1});2^{k-2};)\cdots;2);0)+R(\underbrace{-1,\ldots,-1}_{2^{k}};1).\end{multline}
 Using again recursively the property \ref{pro} the equality (2)
is rewritten as 

\begin{multline}
(\sigma_{(1,2^{k+1})},\sigma_{(2,2^{k+1})},\ldots,\sigma_{(2^{k+1},2^{k+1})})=\\
=R(\underbrace{R(\cdots R(R}_{k-1}(2,2(2^{k-1}+1);2^{k-1});2^{k-2};)\cdots;2);0)+R(\underbrace{R(\cdots R(R}_{k-1}(-1,-1;0);0)\cdots;0);1)=\\
=R(\underbrace{R(\cdots R(R}_{k-1}(2-1,2(2^{k-1}+1)-1;2^{k-1});2^{k-2};)\cdots;2);1)=\\
=\partial R(1,2^{k}+1;2^{k-1})\end{multline}

\end{itemize}
\end{proof}
\begin{ex}\label{seg} In the example \ref{prim}, {\Large {$\sigma_{2^{3}}$}}
has been obtained from {\Large {$\sigma_{2^{2}},$}} both of them
corresponding respectively to the $2^{3}$-periodic orbit and to the
$2^{2}$-periodic orbit. With the theorem ~\ref{fuerte} it is not
necessary to know information about previous orbits to obtain the
results desired, let us see it in the following example.

Applying the theorem~\ref{fuerte}, it results in \[
\text{\Large{$\sigma_{2^{3}}$}}=\partial R(1,2^{2}+1;2^{1})=\partial R(1,5;2).\]
 Using definition \ref{def5}, it is obtained \[
\partial R(1,5;2)=R(R(1,5;2);2^{0})=R(1,5,7,3,;1)=(1\ 5\ 7\ 3\ 4\ 8\ 6\ 2)\]
 that coincides with the obtained in the example \ref{prim} \end{ex}
The theorem ~\ref{fuerte} has the following reformulation 

\begin{thm}
\label{refu} Let $f:\mathrm{I}\to\mathrm{I}$ be an unimodal function
undergoing a period doubling cascade and let $O=\{ x_{1},x_{2},\ldots,x_{2^{p}}\}$
be the $2^{p}$-periodic orbit of the cascade with cardinals $\{ C_{(1,2^{p})}^{*},\ldots,C_{(2^{p},2^{p})}^{*}\}$.
Let the permutation $\text{{\Large{$\sigma_{2^{p}}$}}}=(\sigma_{(1,2^{p})},\,\sigma_{(2,2^{p})},\ldots,\sigma_{(2^{p},2^{p})})\quad p\in\N,\quad p\geq2$
be determined by\[
\text{{\Large{$\sigma_{2^{p}}$}}}=(\sigma_{(1,2^{p})},\,\ldots,\sigma_{(2^{p},2^{p})})=\partial R(1,2^{p-1}+1;2^{p-2})\quad p\in\N,\ p\geq2,\]
 then

\[
C_{(i,2^{p})}^{*}=x_{\sigma_{(i,2^{p})}}\]

\end{thm}
\begin{proof}
By definition $f^{\sigma_{(i,2^{p})}}(C)=x_{\sigma_{(i,2^{p})}}$
and by theorem \ref{fuerte} we get \[
f^{\sigma_{(i,2^{p})}}(C)=C_{(i,2^{p})}^{*}\qquad\text{ with }\ \text{{\Large{$\sigma_{2^{p}}$}}}=\partial R(1,2^{p-1}+1;2^{p-2})\quad p\in\N,\ p\geq2\]
 then \[
C_{(i,2^{p})}^{*}=x_{\sigma_{(i,2^{p})}}\qquad\text{ with }\ \text{{\Large{$\sigma_{2^{p}}$}}}=\partial R(1,2^{p-1}+1;2^{p-2})\quad p\in\N,\ p\geq2.\]
 
\end{proof}
\begin{cor}
\label{cor1} With the conditions of the theorem \ref{refu}, for
every $p\geq2$ with $p\in\N$ the following holds 

\[
f^{i}(C)=C_{(\sigma_{(i,2^{p})}^{-1},2^{p})}^{*}\quad i=1,2,\ldots,2^{p}\]
 where $\text{{\Large{$\sigma_{2^{p}}^{-1}$}}}=(\sigma_{(1,2^{p})}^{-1},\,\sigma_{(2,2^{p})}^{-1},\ldots,\sigma_{(2^{p},2^{p})}^{-1}),$
is the inverse permutation of $\text{{\Large{$\sigma_{2^{p}}$}}}=(\sigma_{(1,2^{p})},\,\sigma_{(2,2^{p})},\ldots,\sigma_{(2^{p},2^{p})})$.
\end{cor}
\begin{proof}
By the Theorem \ref{fuerte} it follows $f^{\sigma_{(j,2^{p})}}(C)=C_{(j,2^{p})}^{*}$
with {\Large {$\sigma_{2^{p}}$}}$=\partial R(1,2^{p-1}+1;2^{p-2})$.
If it is taken $j=\sigma_{(i,2^{p})}^{-1}$ then by definition of
inverse permutation it results in \[
\sigma_{(j,2^{p})}=\sigma_{(\sigma_{(i,2^{p})}^{-1},2^{p})}=i\]
 and therefore $f^{i}(C)=C_{(\sigma_{(i,2^{p})}^{-1},2^{p})}^{*}$
with {\Large $\sigma_{2^{p}}^{-1}$} the inverse of {\Large {$\sigma_{2^{p}}$}}{\Large \par}
\end{proof}
\begin{ex}\label{ter} Taking in mind the results from example \ref{prim},
the inverse permutation of {\Large {$\sigma_{2^{3}}$}} is \[
\text{\Large{$\sigma_{2^{3}}^{-1}$}}=\left(\begin{array}{cccccccc}
1 & 2 & 3 & 4 & 5 & 6 & 7 & 8\\
1 & 8 & 4 & 5 & 2 & 7 & 3 & 6\end{array}\right)\]
that determines in which order are reached the cardinals as the orbit
is visited; that is to say, those are reached in the following order
(see fig. \ref{fig:2a-and-2b}a)

\begin{multline*}
\underbrace{C_{(\sigma_{(1,2^{3})}^{-1},2^{3})}^{*}}_{1}\longrightarrow\underbrace{C_{(\sigma_{(2,2^{3})}^{-1},2^{3})}^{*}}_{8}\longrightarrow\underbrace{C_{(\sigma_{(3,2^{3})}^{-1},2^{3})}^{*}}_{4}\longrightarrow\underbrace{C_{(\sigma_{(4,2^{3})}^{-1},2^{3})}^{*}}_{5}\longrightarrow\underbrace{C_{(\sigma_{(5,2^{3})}^{-1},2^{3})}^{*}}_{2}\longrightarrow\\
\longrightarrow\underbrace{C_{(\sigma_{(6,2^{3})}^{-1},2^{3})}^{*}}_{7}\longrightarrow\underbrace{C_{(\sigma_{(7,2^{3})}^{-1},2^{3})}^{*}}_{3}\longrightarrow\underbrace{C_{(\sigma_{(2^{3},2^{3})}^{-1},2^{3})}^{*}}_{6}\longrightarrow\underbrace{C_{(\sigma_{(1,2^{3}),2^{3})}^{-1}}^{*}}_{1}\end{multline*}

That is to say, the corollary \ref{cor1} indicates that the inverse
permutation determines the order in which the orbit is visited. \end{ex} 

\begin{cor}
\label{sec} With conditions of theorem \ref{refu}, the symbolic
sequence of the orbit is \[
(\mathrm{I}_{(\sigma_{(1,2^{p})}^{-1},2^{p})},\,\mathrm{I}_{(\sigma_{(2,2^{p})}^{-1},2^{p})},\ldots,\mathrm{I}_{(\sigma_{(2^{p},2^{p})}^{-1},2^{p})})\]
 with \[
\left\{ \begin{array}{ll}
\mathrm{I}_{(\sigma_{(2^{p},2^{p})}^{-1},2^{p})}=C\\
\mathrm{I}_{(\sigma_{(i,2^{p})}^{-1},2^{p})}=\left\{ \begin{array}{ll}
\mathrm{R} & \mbox{si $C_{(\sigma_{(i,2^{p})}^{-1},2^{p})}^{*}$ is to the right of $C_{(\sigma_{(2^{p},2^{p})}^{-1},2^{p})}^{*}$ }\\
\mathrm{L} & \mbox{otherwise.}\end{array}\right.\end{array}\right.\]
 
\end{cor}
\begin{proof}
In the orbit with cardinals $\{ C_{(\sigma_{(1,2^{p})}^{-1},2^{p})}^{*},\ldots,C_{(\sigma_{(2^{p},2^{p})}^{-1},2^{p})}^{*}\},$
as $C=C_{(\sigma_{(2^{p},2^{p})}^{-1},2^{p})}^{*}$, we mark $C_{(\sigma_{(2^{p},2^{p})}^{-1},2^{p})}^{*}$
as symbol $'\mathrm{C}'$. Futhermore, if \, $C_{(\sigma_{(i,2^{p})}^{-1},2^{p})}^{*}$
is to the right (left) of \, $C_{(\sigma_{(2^{p},2^{p})}^{-1},2^{p})}^{*}$
we mark it with $\mathrm{R}$\, ($\mathrm{L}$) and the symbolic
sequence of the orbit is thus obtained. 
\end{proof}
\begin{ex} Since $C=C_{(\sigma_{(2^{3},2^{3})}^{-1},2^{3})}^{*}=C_{(6,2^{3})}^{*}$
then $C_{(7,2^{3})}^{*}$ and $C_{(2^{3},2^{3})}^{*}$ are to the
left of $C_{(6,2^{3})}^{*}$ and they are assigned symbol $\mathrm{L}$
according to corollary \ref{sec}. On the other hand $C_{(1,2^{3})}^{*},\ldots,C_{(5,2^{3})}^{*}$
are to the right of $C_{(6,2^{3})}^{*}$ and they are assigned $\mathrm{R}$
according to corollary \ref{sec}. As the inverse permutation, according
example \ref{ter}, is \[
C_{(1,2^{3})}^{*},\ C_{(8,2^{3})}^{*},\ C_{(4,2^{3})}^{*},\ C_{(5,2^{3})}^{*},\
C_{(2,2^{3})}^{*},\ C_{(7,2^{3})}^{*},\ C_{(3,2^{3})}^{*},\ C_{(6,2^{3})}^{*}\]
doing the substitution results in $\mathrm{R}$ $\mathrm{L}$ $\mathrm{R}$
$\mathrm{R}$ $\mathrm{R}$ $\mathrm{L}$ $\mathrm{R}$ $\mathrm{C}$
or, equivalently, $\mathrm{C}$ $\mathrm{R}$ $\mathrm{L}$ $\mathrm{R}$
$\mathrm{R}$ $\mathrm{R}$ $\mathrm{L}$ $\mathrm{R}$.\end{ex} 

The corollary \ref{sec} has the following reformulation:

\begin{cor}
With the conditions of theorem \ref{refu}, \[
C_{(\sigma_{(i,2^{p})}^{-1},2^{p})}^{*}=x_{i}\]

\end{cor}
\begin{proof}
By theorem \ref{refu} it is known that $C_{(j,2^{p})}^{*}=x_{\sigma_{(j,2^{p})}}.$
\, Then, if we take $j=\sigma_{(i,2^{p})}^{-1}$ then \[
C_{(\sigma_{(i,2^{p})}^{-1},2^{p})}^{*}=x_{(\sigma_{(\sigma_{(i,2^{p})}^{-1},2^{p})})}=x_{i}\]

\end{proof}

\section{Conclusions}

Theorem \ref{fuerte} determines the cardinal ordering of fixed points
for any orbit of a period doubling cascade through its associated
permutation {\Large $\sigma_{2^{p}}$}. This result is reached without
using previous information, hence the importance of this result. In
the corollaries that follow from this theorem it is shown, for instance,
that the symbolic sequence of any $2^{p}$-periodic orbit is obtained
from {\Large $\sigma_{2^{p}}$} just giving the number of the bifurcation
$p$. 

Let us point out that topological analysis, in particular, templates,
is a powerful tool to study dynamical systems \cite{Gil98}. Characterization
by templates of dynamical systems is not limited to discrete systems
\cite{lete94}. Symbolic dynamics plays an important role in these
techniques. With the complete characterization presented in this paper,
we complete the information about the dynamics of the system.

\end{document}